\documentclass[%
 amsmath,amssymb,
 aps,onecolumn
]{revtex4-1}

\usepackage{graphicx}
\usepackage{dcolumn}
\usepackage{bm}
\usepackage{algorithm}  
\usepackage{algpseudocode}

\begin{document}

\preprint{APS/123-QED}

\title{Polarization basis tracking scheme for quantum key distribution with revealed sifted key bits}

\author{Yu-Yang Ding$^{1,2}$, Wei Chen$^{1,2,*}$, Hua Chen$^{3}$, Chao Wang$^{1,2}$, Ya-Ping li$^{1,2}$, Zhen-Qiang Yin$^{1,2}$, Shuang Wang$^{1,2}$, Guang-Can Guo$^{1,2}$, Zheng-Fu Han$^{1,2}$}

\affiliation{$^1$ Key Laboratory of Quantum Information, CAS, University of Science and Technology of China, Hefei 230026, China\\
	$^2$ Synergetic Innovation Center of Quantum Information and Quantum Physics, University of Science and Technology of China, Hefei, Anhui 230026, China\\ $^3$  China Aerospace Science and Technology Corporation\\
	\noindent{$^*$:Corresponding author: kooky@mail.ustc.edu.cn}
}

\date{\today}

\begin{abstract}
Calibration of the polarization basis between the transmitter and receiver is an important task in quantum key distribution (QKD). An effective polarization-basis tracking scheme will decrease the quantum bit error rate (QBER) and improve the efficiency of a polarization encoding QKD system. In this paper, we proposed a polarization-basis tracking scheme using only unveiled sifted key bits while performing error correction by legitimate users, rather than introducing additional reference light or interrupting the transmission of quantum signals. A polarization-encoding fiber BB84 QKD prototype was developed to examine the validity of this scheme. An average QBER of 2.32$\%$ and a standard derivation of 0.87$\%$ have been obtained during 24 hours of continuous operation.
\begin{description}
	\item[PACS numbers]
	03.67.Dd, 03.67.Hk, 42.50.Ex
\end{description}
\end{abstract}

\pacs{Valid PACS appear here}
\maketitle


Quantum key distribution (QKD) could in principle implement information-theoretic security (ITS)  key reconciliation using quantum mechanics. Since Bennett and Brassard proposed the most famous QKD protocol (BB84) in 1984\cite{bennett2014quantum}, further advances have been made both in theory and experiment \cite{lo2014secure,scarani2009security,gisin2002quantum}. In most QKD experiments, photons are natural resources to carry quantum information, which can be transferred through fiber or free space channels. Fiber is the most widely used quantum channel profiting from its extensibility, while its birefringence has disturbed experimental researchers for many years. The birefringence of the fiber makes QKD systems vulnerable to environmental disturbance and significantly affects the performance of QKD systems. For example, the polarization reference frames of a transmitter Alice and a receiver Bob will be misaligned in polarization encoding QKD systems, which will directly increase the quantum bit error rate (QBER) of the system\cite{chen2007active}. In phase encoding QKD systems, polarization fluctuation will reduce the fringe visibility and consequentially increase the QBER of the systems.\cite{hiskett2006long,wang2015phase}. Due to its adverse effects, effective polarization compensation schemes should be performed in QKD systems.

Some passive self-compensation methods have been developed for phase encoding QKD systems, such as the "Plug-and-play" \cite{Muller1997plugandplay} and "Faraday-Michelson" \cite{Mo2005FMI} structures, while active feedback tracking is still a major methods used in polarization encoding QKD systems \cite{chen2007active,ma2006polarization,xavier2008full,chen2009stable,lucio2009proof,xavier2009experimental,franson1995operational,de2008polarisation}. Depending on whether it halts the quantum photon transmitting procedure, most of these strategies can be divided into two types, the interrupting scheme \cite{chen2007active,ma2006polarization,lucio2009proof,chen2009stable} and the real-time scheme \cite{xavier2008full,xavier2009experimental,de2008polarisation}. The former will apparently sacrifice the efficiency of the system. The latter is primarily based on the time-division multiplexing (TDM) and the wavelength-division multiplexing (WDM) methods. The TDM methods send the reference signals in the time intervals between quantum photons, which may not be compatible with up-to-date high-speed QKD systems. The WDM methods send reference signals with the wavelength distinct from the quantum signals, thus may limit the security transmission distance of QKD due to the polarization decorrelation between the two signals \cite{chen2009stable,muga2011qber}. Because most of these schemes need extra components such as photodetectors or laserdiodes, the QKD system will be more complicated and will need to be carefully designed to protect the quantum photons from the ``pollution'' of the relatively strong reference light.

In this paper, we propose a real-time continuous polarization-basis tracking scheme for QKD based on single photon detection signals. We use the sifted key bits unveiled and discarded during the post processing procedure of QKD, which are usually $10\%$ to $20\%$ of the total sifted key bits \cite{scarani2009security,poppe2004practical,gisin2002quantum}, to calculate the feedback control signals of a gradient algorithm that are applied to the polarization control (PC) components accordingly. Because the scheme can be implemented without extra photon sources, detectors and time expenditures, the method can be used in some single photon level quantum optical experiments\cite{Yuan2008, Xavier2008}. We applied the scheme to the off-the-shelf electronic polarization controllers (EPCs) in a typical polarization encoding QKD system to compensate for the polarization disturbance of a 50 km fiber channel. The polarization-basis tracking can be performed during the post-processing of QKD without interrupting the transmission of quantum signals. The average QBER of the QKD system is 2.32$\%$ and the standard derivation is 0.87$\%$ during 24 hours of continuous operation, which effectively verified the validity of using this method in QKD.

The polarization-basis tracking method should be used in conjunction with specific polarization control components in QKD systems. The EPCs based on fiber-squeezers \cite{yao2002fiber} are fundamental devices used in QKD system profiting from their low insertion loss. Major drawbacks of this type of EPCs are low modulating rates and inconsistency even when the same driving voltages are applied. Thus, an optimizing algorithm should be developed to make them work efficiently. We selected the four-stage EPCs (PolaRITE \uppercase\expandafter{\romannumeral3}, General Photonics, 5228 Edison Avenue
Chino, CA 91710) as the polarization control unit, which has an insertion loss as low as 0.05 dB. The EPC consists of four electronic fiber squeezers (FSs) $X_1$,$X_2$,$X_3$ and $X_4$. The functions of $X_3$ and $X_4$ are the same as $X_1$ and $X_2$, respectively. The voltage applied on one FS will rotate the state of polarization (SOP) with an angle of $\varphi_i$ around the axis $\Omega_i$ ($i=1,2,3,4$) on the $\acute{Poincare}$ sphere. The rotation $\Re_i$ can be described by the Hamilton's $quaternion$\cite{kuipers1999quaternions}: 
\begin{equation}
\Re_i=\cos{\frac{\varphi_i}{2}} + \sin{\frac{\varphi_i}{2}} \Omega_i
\end{equation}
where $\varphi_i$ is determined by the voltage $v_i$ on the FS$_i$, and $\Omega_i$ represents the rotation axis on the $\acute{Poincare}$ sphere. The SOP $\left| S_1\right\rangle $ after rotating an input state $\left| S_0\right\rangle $ with one squeezer can be represented by their normalized Stokes vectors $S_0$ and $S_1$:
\begin{equation}
S_1=\Re_i S_0 \Re_i^{*}
\end{equation}

In principle, any SOP changing in the fiber channels can be compensated by driving the EPC with proper voltages. Three FSs are sufficient to track all polarization states and the algorithm can be made more flexible with extra FSs \cite{walker1990polarization}. The driving voltages are adjusted according to the feedback signal $E$, which is a function of the current values of the driving voltages $v_i$ and $\Omega_i$ (i=1, 2, 3, and 4). Unfortunately, the directions of the FS's rotation axes change over time due to their mechanical structures and environmental disturbances. As a result, there are two challenges to implementing a real-time polarization tracking algorithm with this type of EPCs: $a) $ it is difficult to build a look-up table for each squeezer and $\Omega_i$ must be measured each time before adjustment and, $b)$ the algorithm works during quantum key transmission, which means it is unavailable to traverse all combinations of EPC driving voltages. To overcome these disadvantages, we modified the widely used Gradien algorithm according to the traits of EPC and QKD systems. The flow chart of the algorithm can be described as in algorithm 1.

\begin{figure*}[htbp]
	\centering
	\includegraphics[width=\linewidth]{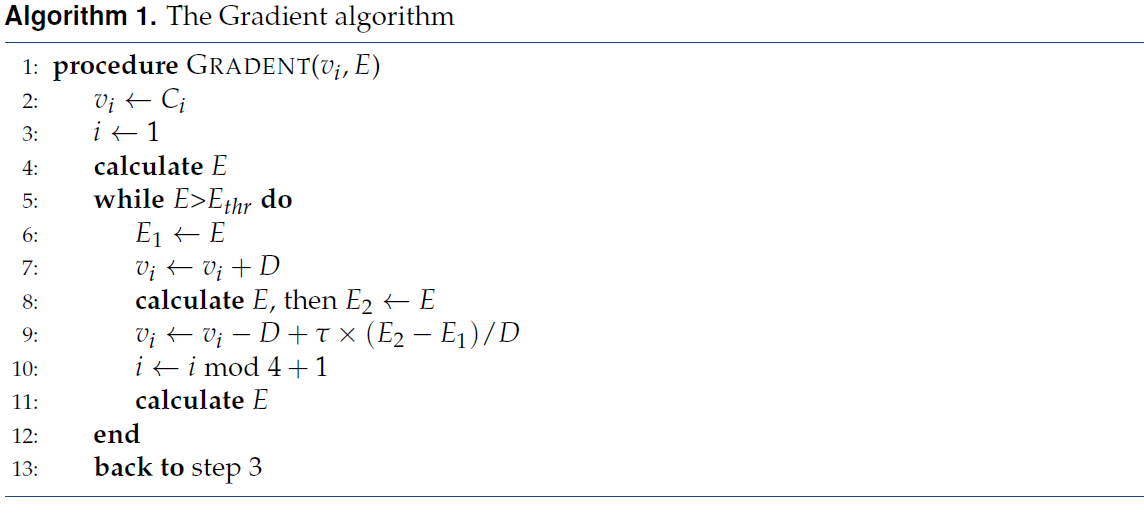}
	\label{fig:EX1 set}
\end{figure*}

In algorithm 1, $i=$1,2,3 and 4; $C_i$ is the center of the nomal range of $v_i$. The arrows $\gets$ represents the assignment operation, $v_i$ is the immediate driving voltage on the EPC squeezer's axis $\Omega_i$ ($i$=1,2,3 and 4), $E$ is the feedback value and $E_{thr}$ is the threshold value of the deviation from the ideal situation in the QKD system. The parameter $D$ is a small dither value and $\tau$ is an amplification factor of the voltage adjustment. At the beginning of the algorithm, the four $v_i~(i=1,2,3,4)$ were initialized as the center value of the dynamic range of the driving voltage. Because the rotation of $\Omega_i$ is slowly varying, it can be regarded as a constant from step 6 to step 11 within a cycle of the algorithm. To estimate the reference value of $\Omega_i$, a tiny dither voltage $D$ is applied to the FS to lightly change $\varphi$ in step 7. The deviation values $E$ before and after fine-tuning are calculated and used to obtain the partial derivative $(E_2-E_1)/D$\cite{chen2016new,noe1988endless}, which shows the relationship between $E(v_i, \Omega_i)$ and $v_i$, in other words, the direction of $\Omega_i$ at this moment. In step 9, we modify $v_i$ according to the partial derivation $(E_2-E_1)/D$, and $D$ is subtracted so that $\Omega_i$ can return back to its original position. $\tau$ is used to tune the step size and is negative to minimize $E$. The four $v_i$ (i=1,2,3 and 4) are adjusted one by one as described in step 10 until $E$ is lower than the threshold $E_{thr}$. The four driving voltages of the FSs are kept until $E$ exceeds $E_{thr}$ again. 

The crux of this algorithm is how to accurately calculate the feedback signal $E$ with single photon signals. In the polarization encoding BB84 protocol, Alice randomly modulates the polarization of photons to horizontal states $\left| H\right\rangle $, vertical states $\left| V\right\rangle $, $45^\circ$ states $\left|H+V\right\rangle$, and $135^\circ$ states $\left|H-V\right\rangle$. The four states are combined to make two conjugate bases, which are usually named the $Z$ basis ($\left| H\right\rangle$ and $\left| V\right\rangle$) and $X$ basis ($\left|H+V\right\rangle$ and $\left|H-V\right\rangle$). Bob randomly selects one of the two bases to measure the photons. The tracking scheme for the two basis is identical, so that we use the $Z$ basis to illustrate the estimation method of $E$. Comparing the states sent by Alice and detected by Bob, we can obtain an MM $U$, which can be depicted as follows:
\begin{equation}
U=\bordermatrix{
	Alice\backslash Bob & \left| H\right\rangle & \left| V\right\rangle \cr
	\left| H\right\rangle & j_1 & j_2 \cr
	\left| V\right\rangle & j_3 & j_4 }
\label{eq:UI}
\end{equation}
where $j_1$ and $j_2$ ($j_3$ and $j_4$) denote the probabilities that Bob gets $\left| H\right\rangle $ and $\left| V\right\rangle $ when Alice sends $\left| H\right\rangle $ ($\left| V\right\rangle $), respectively.Nomalization requires that,  $j_1+j_2=1$ and $j_3+j_4=1$ for matrix $U$. Ideally, when Alice sends a $\left| H\right\rangle$ state photon, Bob will obtain a deterministic result of $\left| H\right\rangle$ when he measures the photon with the $Z$ basis. It is similar for the other three states. Thus, a reference matrix (RM) in the ideal situation should be an identity matrix $I$. However, a practical QKD system will suffers from the environmental disturbance and its intrinsic noise. As a result, the measurement matrix (MM) of single photon detection results will deviate from the RM. Thus, the distance of $U$ and $I$ during quantum key transmission will be highly related to the rotation degree of Alice and Bob's polarization reference frame and can be used to calculate the feedback signal $E$. The method we used to calculate the distance between RM and MM is described in the following equation.

\begin{equation}
E =\sum_i\sum_j(U_{ij}-I_{ij})^2
=2({j_2}^2+{j_3}^2) 
\label{eq:E}
\end{equation}

The sample size required to perform the algorithm is an important parameter, because more samples means a longer acquisition time, which will weaken the tracking capability. The final purpose of polarization-basis tracking is to minimize the QBER, which is defined as the ratio between the wrong and the total detection results when Alice and Bob select the same basis \cite{muga2011optimization}. We divide the QBER into three parts as follows:
\begin{equation}
QBER = QBER_b+QBER_d= \frac{K_{wrong}}{K_{shift}}
\label{eq:QBER}
\end{equation}
where $QBER_b$ is caused by the reference frame mismatch between Alice and Bob, which we need to compensate during the QKD procedure, $QBER_d$ is caused by the intrinsic imperfections of the QKD system, such as the modulating errors and the dark counts of the single photon detectors (SPD) which is approximately 1$\%$ to 1.5$\%$ in an typical system below. $K_{wrong}$ represents the rate of the error key after Alice and Bob compare their measurement basis, and $K_{shift}$ is the rate of the shifted key.

\begin{figure}[htbp]
	\centering
	\includegraphics[width=\linewidth]{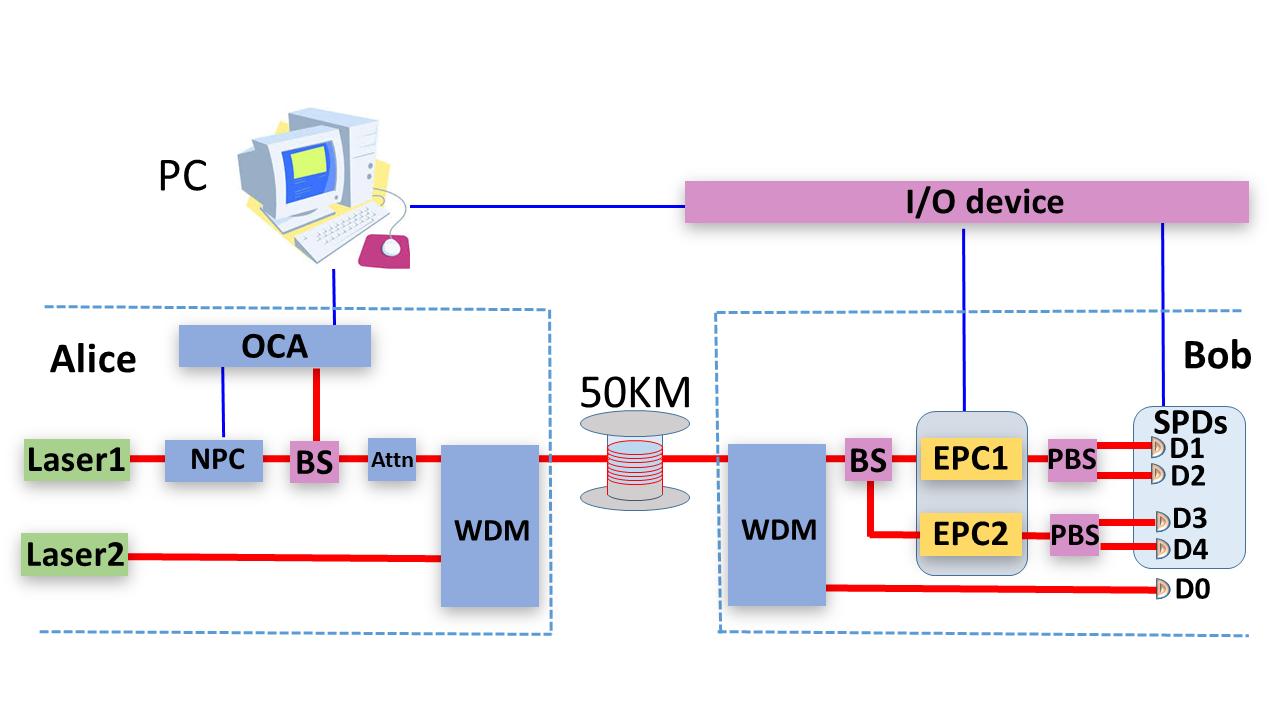}
	\caption{(Color online) Schematic of the polarization compensation experiment set up. Thick lines (red online) represent the light path and thin lines (green online) represent electric circuits. Laser1, 1550.92 nm DFB pulse laser for quantum light; Laser2, 1549.32 nm DFB pulse laser for synchronous light; NPC,$LiNbO_3$ (lithium niobate) polarization control; BS, (50:50) beam splitter; ATTn, attenuator; EPC1, EPC2 and EPC3, electronic polarization controller; PBS, polarization beam splitter; D1,D2,D3 and D4, single photon detectors; OCA, optical component analyzer; WDM: wavelength division multiplexer; D$_0$, the synchronous detector; PC, personal computer. }
	\label{fig:EX1 setup}
\end{figure}

We present an experiment to verify the effectiveness of our algorithm in a typical BB84 polarization encode QKD system, the system is shown schematically in Fig. \ref{fig:EX1 setup}. Alice sends Bob a sequence of photons prepared in different states, which are chosen randomly from two conjugate polarization bases, a rectilinear basis and a diagonal basis. We use a $LiNbO_3$ (lithium niobate) polarization controller to achieve the state-selection process \cite{hidayat2008fast,xavier2009experimental}, with a repetition rate of 2.5 MHz. The optical component analyzer (OCA) is used to monitor the states that Alice prepares. The mean photon numbers of the laser at Alice is attenuated to 0.1 per pulse. There are two electronic polarization controllers (EPC) on Bob's side, EPC$_1$ and EPC$_2$, for aligning the reference frame for each measuring basis separately between Alice and Bob. Each SPD in our experiment has s detection efficiency of approximately 10$\%$.

The finite number of samples will affect the accuracy of the QBER estimation and polarization-basis tracking. A larger block size means minor deviation and a longer acquisition time. We analyzed the relationship between the sample size, the QBER, and the estimation error (discussed in detail in the supplemental materials). In our experiment, $10\%$ of the sifted key bits(approximately 2500 bits) revealed during the post-possessing \cite{scarani2009security,poppe2004practical,gisin2002quantum} are used to estimate the polarization basis in each feedback loop, and an estimation error of less than $0.6\%$ has been obtained when the $QBER$ is below $10\%$. 

\begin{figure*}[!htb]
	\centering
	\includegraphics[width=\linewidth]{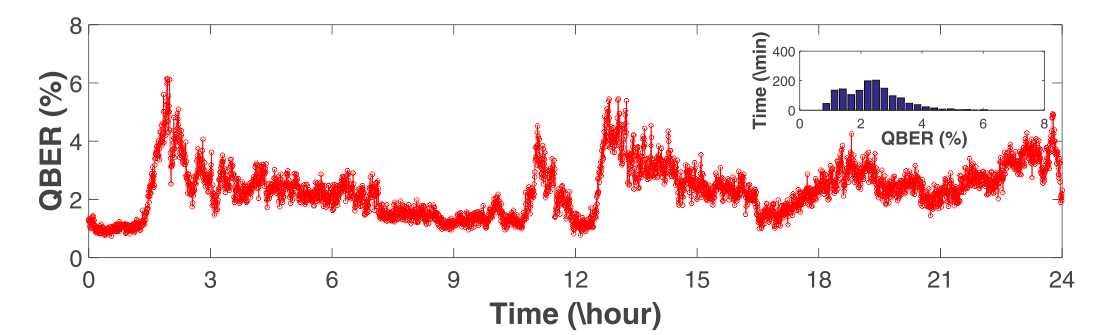}
	\caption{(Color online) QBER of polarization-coding QKD using the polarization control strategy working during 24 hours. The inset histogram is the distribution of QBER over time.}
	\label{fig:test2}
\end{figure*}

The Fig. \ref{fig:test2} is the QBER of whole system during the 24 hours test time. When the polarization control is working, the QBER for the QKD system remains at a very low level in most of the test time. The average QBER is down to 2.32$\%$ and the standard derivation of QBER is 0.87$\%$. For the real life QKD systems, in which polarization in the fiber changes slowly for most of the transit time, our strategy is effective for polarization control. Also we perform another experiment to test the feasibility of our scheme in a fast scrambling situation. 

\begin{figure*}[!htbp]
	\centering
	\includegraphics[width=\linewidth]{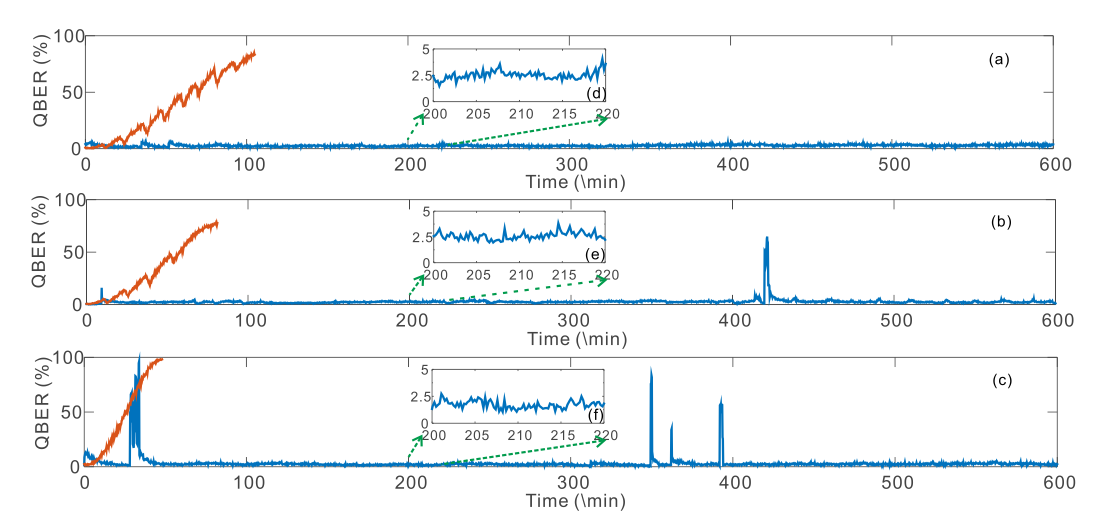}
	\caption{(Color online) With (blue lines) and without (red lines) polarization control when dither rotation speeds are $0.2^{\circ}$ (a), $0.4^{\circ}$ (b), $0.6^{\circ}$(c) per FC. Insets (d), (e), (f) detail (200 min to 220 min) of (a), (b), (c) separately.}
	\label{fig:test1}
\end{figure*}

In the polarization scrambling test, we replace 50 km of telecom fiber with another EPC$_3$ (not show in Fig. \ref{fig:EX1 setup}). The EPC$_3$ is used to generate a slowly-varying polarization change by modifying the voltage applied to one of its fiber squeezers \cite{shimizu1991highly}. The polarization state will rotate around an axis on the $\acute{Poincare}$ sphere from the state that Alice prepared to an unknown state $\left| \psi \right\rangle $ because of the scrambling of the EPC$_3$. Fig. \ref{fig:test1} shows three tests under different rotation speeds, namely $0.2^{\circ}$, $0.4^{\circ}$, $0.6^{\circ}$ per FC (12 s in our experiment). The blue lines depict the measurements with use of the control scheme and the results without using the control scheme are depicted by red lines as comparison. Considering the periodicity of the scrambling, we erase the repeating part of the red lines and magnify parts of the blue line (200 min to 220 min) in (d) , (e) and (f) for higher clarity. In three tests, when the polarization control is working, the average QBER during 10 hours are 2.65$\%$, 2.74$\%$ and 3.29$\%$ for (a), (b) and (c). For figure (a), compared to red line, the green line maintains a very low level during the whole test time. When the scrambling angle for EPC$_3$ becomes bigger in (b) and (c), the QBER is steady for most of time, but there are few abrupt rises of QBER. Actually, the tracking ability of our control system is strongly limited by the long FC, when the polarization change in the fibers is abouptly-varying, our strategy might not obtain the feedback information in time. However, such events of rising rarely occur. By increasing our low system frequency (2.5 MHz), the FC can be reduced significantly and, then the tracking performance of our control scheme will be greatly enhanced\cite{chen2016new}. 

In conclusion, we have proposed and experimentally demonstrated an effective single-photon level polarization tracking scheme. The feedback control parameters of the scheme are calculated using revealed and discarded sifted bits during the  post-processing procedure of QKD sessions, which makes it a real-time, effective method. Although the experiment is based on the BB84 protocol, the scheme is suitable for use in various QKD systems, such as phase-encoding QKD systems and measurement-device-independent QKD systems. In our present work, the working frequency of the QKD system limits the tracking capability. Considering that the working frequency of QKD systems has previously exceeded 1GHz\cite{comandar2016quantum}, the tracking capability of the scheme has great potential to be significantly improved\cite{chen2016new}.

This work has been supported by the National Basic Research Program of China (Grants No. 2011CBA00200 and No. 2011CB921200), the National Natural Science Foundation of China (Grant Nos. 61475148, 61575183), and the ``Strategic Priority Research Program (B)" of the Chinese Academy of Sciences (Grant Nos. XDB01030100, XDB01030300).

\section{Supplemental Material}

For our scheme, using a large enough sample size is important when tracking the polarization state, because the statistical fluctuation of the samples will weaken the reference tracking ability. Determining the required unmber of  keys in a sample is a considerable problem. The photons, before going through the PBS with an arbitrary polarization, (we use the rectilinear base in Fig. \ref{fig:EX1 setup} as an example), can be described by \cite{muga2011qber}

\begin{equation} 
\left| \phi\right\rangle = \sin \theta \left| V\right\rangle + \cos \theta e^{i \phi} \left| H\right\rangle
\end{equation}
where $\left| V\right\rangle$ and $\left| H\right\rangle$ represent the states at the exit through the vertical and horizontal ports for PBS, respectively. $\theta$ is related to the polarization projection, and $\phi$ is the retardation between $ \left| H\right\rangle$ and $\left| V\right\rangle$. Then, we use $A_1$, $A_2$ to represent the fraction of photons that exit through the PBS. 

\begin{equation} 
A_1=\left|\left\langle V| \phi \right\rangle  \right| ^2= 1-\cos^2 \theta
\end{equation}

\begin{equation} 
A_2=\left|\left\langle H| \phi\right\rangle \right| ^2= \cos^2 \theta
\end{equation}
so the probability for detector $D1$ and $D2$ to obtain counts is\cite{ma2005practical}

\begin{equation}
P_1=1-e^{-\eta\left\langle n_1\right\rangle }
\label{1}
\end{equation}
\begin{equation}
P_2=1-e^{-\eta\left\langle n_2\right\rangle }
\label{2}
\end{equation}
where  $\left\langle n_1\right\rangle = A_1 \left\langle n\right\rangle$ and $\left\langle n_2\right\rangle = A_2 \left\langle n\right\rangle$ are the mean number of data photons per pulse to $D1$ and $D2$ after PBS. $\left\langle n\right\rangle$ is the mean photons number per pulse for light launched by Alice. $\eta$ is the overall transmission and detection efficiency between Alice and Bob

\begin{equation} \label{eta}
\eta=t_{ab}\eta_{Bob}=10^{-\alpha l/10}\eta_{Bob}
\end{equation}

For simplicity, we suppose the devices in our experiment are perfect, so we ignore the influence of the dark counts in the SPDs there. Where $\alpha$, $l$ and $\eta_{Bob}$ are the loss coefficient in dB/Km, the length of fiber in Km and the detection efficiency of Bob's detectors, respectively. Without loss of generality, we suppose Alice only sends one polarization state $\left| H\right\rangle$. We use $QBER_t$ to represent the true error rite in the fiber. Then, we have

\begin{equation}
QBER_t = \left\langle n_1\right\rangle/ (\left\langle n_1\right\rangle+\left\langle n_2\right\rangle)= 1-\cos^2 \theta
\end{equation}

In our scheme, we use $QBER_e$ to estimate $QBER_t$ which is used as feedback by consuming shift keys. The sample size of consumed shift keys is $B$, and the counts for $D1$ and $D2$ are $M$ and $N$ after Alice sends $B$ pulses to Bob. Then

\begin{equation}\label{6}
QBER_e = M/(M+N) =M/B
\end{equation}
$M$ and $N$ approach the Binomial distribution

\begin{equation}
P(M=m)=\binom{B}{m}P_1^m (1-P_1)^{Q-m}
\label{3}
\end{equation}

\begin{equation}
P(N=n)=\binom{B}{n}P_2^n (1-P_2)^{Q-n}
\label{4}
\end{equation}

If the sample size is infinite, we have $QBER_e = QBER_t$, but for real circumstances, there is a deviation between $QBER_e$ and $QBER_t$. According to probability theory, we know there is a $99.7\%$ probability for $QBER_e$ to be in the section $[E(QBER_e)-3*\sigma_{QBER_e} \qquad E(QBER_t)+3*\sigma_{QBER_e}]$, where $E(QBER_e)$ and $\sigma_{QBER_e}$ are the expectation and standard deviation of $QBER_e$. Because $E(QBER_e) = QBER_t$ obviously, we use $\Delta_{QBER}=3*\sigma_{QBER_e}$ as the maximum offset for $QBER_e$ from $QBER_t$. According to Eq.\ref{6} and Ref.\cite{johnson1999survival,struart1999kendall}, we have

\begin{equation}\begin{split}\label{5} 
\sigma_{QBER_e} &=\sqrt{Var(\frac{M}{B})}\\
&\approx \sqrt{\frac{1}{E^2(Q')}Var(M) - 2\frac{E(M)}{E^3(Q')}Cov(M,B) + \frac{E^2(M)}{E^4(Q')}Var(B)}\\
\end{split}\end{equation}
where $Var$ means variance, and $Cov$ represent covariance. By substituting Eqs.\ref{3},\ref{4} and \ref{5}, we get

\begin{equation}\label{8}
\Delta_{QBER} = 3*\sigma_{QBER_e} \approx 3*\sqrt{\frac{1}{B}*\frac{P_1 P_2 (P_1+P_2-2P_1P_2)}{(P_1+P_2)^3}}
\end{equation}

\begin{figure}[htbp]
	\centering
	\includegraphics[width=0.8\textwidth]{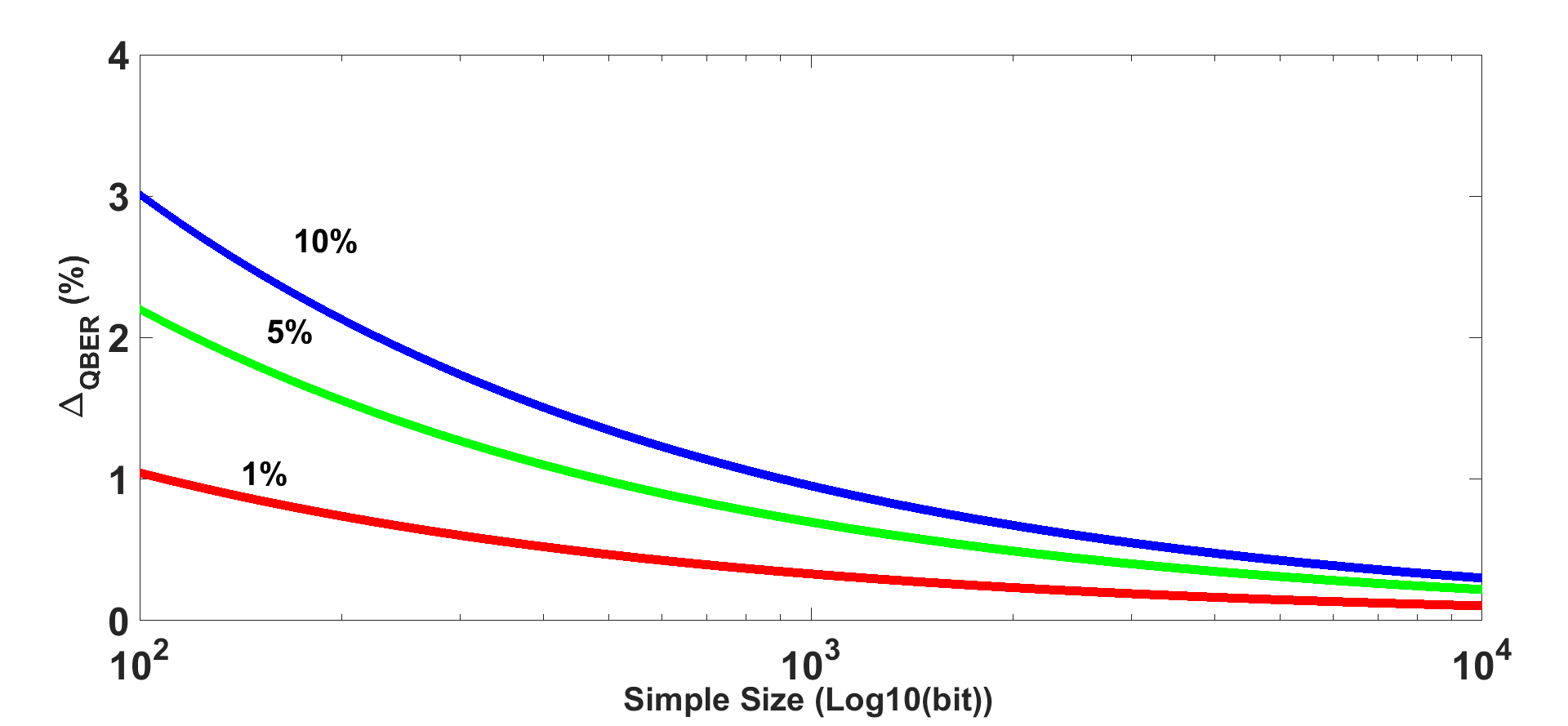}
	\caption{(Color online) the relationship of the consumed bits used to track the SOP and the standard deviation for $QBER_e$ given different $QBER_t$.}
	\label{fig:QBER}
\end{figure}

The maximum offset $\Delta_{QBER}$ plotted versus sample size is show in Fig. \ref{fig:QBER}. Consider Eqs\ref{1}, \ref{2}, \ref{8}, with $QBER=1\%, 2\%, 3\%$ respectively and with $\eta=10\%$, $\left\langle n\right\rangle=0.1$ in all the case.

According the analysis above, and considering the low repetition rate of our QKD system, the sample size is chosen to be 2500 bits for one polarization base. Under these circumstances, $\Delta_{QBER}$ is under 0.6$\%$ when $QBER_t$ below $10\%$. If $QBER_t$ is greater than $10\%$, we use all the sifted keys to track the polarization reference disalignment, and it is enough for our scheme obviously. Of course, when the repetition rate rises, we can sample more bits for polarization tracking, and $\Delta_{QBER}$ will obviously be less .                                                                                                                                                                                                                                                                                                                                                           




\end{document}